\documentclass[12pt]{article}
\usepackage{url}
\usepackage{graphics,epsfig}
\usepackage{amsmath,amssymb,amsthm}
\usepackage{graphicx}
\usepackage{epstopdf}
\def\directunion{\hbox{$\bigcirc$ \hskip - 11.3 pt \raise 0.1pt
\hbox{$\scriptstyle \vee$}}\ }
\def\orthrel{\hbox{$\perp$ \hskip - 10.3 pt \raise 3.2pt \hbox{$\scriptstyle \wedge$}}\ }
\newtheorem{definition}{Definition}

\newtheorem{proposition}{Proposition}

\newcommand{\bibnodot}[1]{}

\newcommand{\real}{{\mathbb R}}

\newcommand{\bd}{\begin{definition}}
\newcommand{\ed}{\end{definition}}
\newcommand{\bp}{\begin{proposition}}

\title{Operational Quantum Mechanics, Quantum Axiomatics and Quantum Structures\footnote{Published as: Aerts, D. (2009). Operational quantum mechanics, quantum axiomatics and quantum structures. In D. Greenberger, K. Hentschel and F. Weinert, (Eds.), {\it Compendium of Quantum Physics: Concepts, Experiments, History and Philosophy}. New York: Springer.}}
\author{Diederik Aerts}
\date{}
\begin{document}

\maketitle
\centerline{Leo Apostel Center (CLEA) and Department of Mathematics (FUND),}
\centerline{Brussels Free University, Krijgskundestraat 33, 1160 Brussels, Belgium.}
\centerline{E-mail: \url{diraerts@vub.ac.be}} 

\bigskip
\bigskip
\noindent
Operational quantum mechanics and quantum axiomatics have their roots in a work of John von Neumann in collaboration with Garett Birkhoff, that is almost
as old as quantum mechanics itself [1]. Indeed already during the
beginning years of quantum mechanics, the formalism that is now referred to as standard quantum mechanics
[2], was thought to be too specific by the founding fathers themselves. One of the questions that obviously
was at the origin of this early dissatisfaction is: `Why would a complex Hilbert space deliver the unique mathematical
structure for a complete description of the microworld? Would that not be amazing? What is so special about a complex
Hilbert space that its mathematical structure would play such a fundamental role?'

Let us turn for a moment to the other great theory of physics, namely general relativity, to raise more suspicion towards the fundamental
role of the complex Hilbert space for quantum mechanics. General relativity is founded on the mathematical structure of Riemann geometry. In this
case however it is much more plausible that indeed the right fundamental mathematical structure has been taken. Riemann developed his theory as a
synthesis of the work of Gauss, Lobatsjevski and Bolyai on non-Euclidean geometry, and his aim was to work out a theory for the description of the
geometrical structure of the world in all its generality. Hence Einstein took recourse to the work of Riemann to express his ideas and
intuitions on space time and its geometry and this lead to general relativity. General relativity could be called in this respect `the
geometrization of a part of the world including gravitation'.

There is, of course, a definite reason why von Neumann used the mathematical structure of a complex Hilbert space for the formalization
of quantum mechanics, but this reason is much less profound than it is for Riemann geometry and general relativity. The reason is that
Heisenberg's matrix mechanics and Schr\"odinger's wave mechanics turned out to be equivalent, the first being a formalization of the new
mechanics making use of
$l_2$, the set of all square summable complex sequences, and the second making use of $L_2(\real^3)$,
the set of all square integrable complex functions of three real variables. The two spaces $l_2$ and $L_2(\real^3)$ are canonical examples
of a complex Hilbert space. This means that Heisenberg and Schr\"odinger were working already in a complex Hilbert space, when they
formulated matrix mechanics and wave mechanics, without being aware of it. This made it a straightforward choice for von Neumann to
propose a formulation of quantum mechanics in an abstract complex Hilbert space, reducing matrix mechanics and wave mechanics to two
possible specific representations.

One problem with the Hilbert space representation was known from the start. A (pure) state of a quantum entity is represented by a
unit vector or ray of the complex Hilbert space, and not by a vector. Indeed vectors contained in the same ray represent the same
state or one has to renormalize the vector that represents the state after it has been changed in one way or another. It is well known
that if rays of a vector space are called points and two dimensional subspaces of this vector space are called lines, the set of points and lines
corresponding in this way to a vector space, form a projective geometry. What we just remarked about the unit vector or ray representing the
state of the quantum entity means that in some way the projective geometry corresponding to the complex Hilbert space represents more
intrinsically the physics of the quantum world as does the Hilbert space itself. This state of affairs is revealed explicitly in the dynamics of
quantum entities, that is built by using group representations, and one has to consider projective representations, which are representations in
the corresponding projective geometry, and not vector representations [3].

The title of the article by John von Neumann and Garett Birkhoff [1] that we mentioned as the founding article for operational quantum axiomatics is `The logic of quantum mechanics'. Let us explain shortly what Birkhoff and von Neumann do in this article. First of all they remark
that an operational proposition of a quantum entity is represented in the standard quantum formalism by an orthogonal projection operator or
by the corresponding closed subspace of the Hilbert space ${\cal H}$. Let us denote the set of all closed subspaces of ${\cal H}$ by ${\cal
L}({\cal H})$. Next Birkhoff and von Neumann show that the structure of ${\cal L}({\cal H})$ is not that of a Boolean algebra, the archetypical
structure of the set of propositions in classical logic. More specifically it is the distributive law between conjunction and disjunction 
\begin{equation} \label{eq:distributivity}
(a \vee b) \wedge c = (a \wedge c) \vee (b \wedge c)
\end{equation}
that is
not necessarily valid for the case of quantum propositions $a, b, c \in {\cal L}({\cal H})$. A whole line of research, called
quantum logic, was born as a consequence of the Birkhoff and von Neumann article. The underlying philosophical idea is that, in the same manner
as general relativity has introduced non-Euclidean geometry into the reality of the physical world, quantum mechanics introduces non-Boolean
logic. The quantum paradoxes would be due to the fact that we reason with Boolean logic about situations with quantum entities, while these
situations should be reasoned about with non-Boolean logic. 

Although fascinating as an approach [4], it is not this
idea that is at the origin of quantum axiomatics. Another aspect of what Birkhoff and von Neumann did in their article is that they shifted the attention on the mathematical structure of the set of operational propositions ${\cal L}({\cal H})$ instead of
the Hilbert space
${\cal H}$ itself. In this sense it is important to pay attention to the fact that ${\cal L}({\cal H})$ is the set of all operational
propositions, {\it i.e.} the set of yes/no experiments on a quantum entity. They opened a way to connect abstract mathematical concepts of the
quantum formalism, namely the orthogonal projection operators or closed subspaces of the Hilbert space, directly with physical operations in the
laboratory, namely the yes/no experiments.

George Mackey followed in on this idea when he wrote his book on the mathematical foundations of quantum mechanics
[5]. He starts the other way around and considers as a basis the set ${\cal L}$ of all operational propositions, meaning propositions
being testable by yes/no experiments on a physical entity. Then he introduces as an axiom that this set ${\cal L}$ has to have a structure
isomorphic to the set of all closed subspaces
${\cal L}({\cal H})$ of a complex Hilbert space in the case of a quantum entity. He states that it would be interesting to invent a set of axioms
on ${\cal L}$ that gradually would make
${\cal L}$ more and more alike to ${\cal L}({\cal H})$ to finally arrive at an isomorphism when all the axioms are satisfied. While Mackey wrote
his book results as such were underway. A year later Constantin Piron proved a fundamental representation theorem.
Starting from the set ${\cal L}$ of all operational propositions of a physical entity and introducing five axioms on ${\cal L}$ he proved that
${\cal L}$ is isomorphic to the set of closed subspaces ${\cal L}(V)$ of a generalized Hilbert space $V$ whenever these five axioms are satisfied
[6]. Let us elaborate on some of the aspects of this representation
theorem to be able to explain further what operational quantum axiomatics is about.

We mentioned already that Birkhoff and von Neumann had noticed that the set of closed subspaces ${\cal L}({\cal H})$ of a complex Hilbert space
${\cal H}$ is not a Boolean algebra, because distributivity between conjunction and disjunction, like expressed in (\ref{eq:distributivity}), is not satisfied. The set of closed subspaces of a complex Hilbert space forms however a lattice, which is a more general
mathematical structure than a Boolean algebra, moreover, a lattice where the distributivity rule (\ref{eq:distributivity}) is satisfied
is a Boolean algebra, which indicates that the lattice structure is the one to consider for the quantum mechanical situation. To make again a reference to general relativity, the lattice structure is indeed to a Boolean algebra what general
Riemann geometry is to Euclidean geometry. And moreover, meanwhile it has been understood why the structure of operational propositions of the world
is not a Boolean algebra but a lattice. This is due to the fact that measurements can have an uncontrollable influence on the state of
the physical entity under consideration [7]. Hence
the intuition of Birkhoff and von Neumann, and later Mackey, Piron and others, although only mathematical intuition at that time, was correct.

Axiomatic quantum mechanics is more than just an axiomatization of quantum mechanics. Because of the operational nature of the axiomatization, it holds the potential for `more general theories than standard quantum mechanics' which however are `quantum like theories'. In this sense, we believe that it is one of the candidates to generate the framework for the
new theory to be developed generalizing quantum mechanics and relativity theory [7]. Let us explain why we believe that operational quantum axiomatics has the potential to deliver such a generalization of relativity theory and quantum mechanics. General relativity is a theory that brings
part of the world that in earlier Newtonian mechanics was classified within dynamics to the geometrical realm of reality, and more specifically
confronting us with the pre-scientific and naive realistic vision on space, time, matter and gravitation. It teaches us in a deep and new way,
compared to Newtonian physics, `what are the things that exists and how they exist and are related and how they influence each other'. But there
is one deep lack in relativity theory: it does not take into account the influence of the observer, the effect that the measuring
apparatus has on the thing observed. It does not confront the subject-object problem and its influence on
how reality is. It cannot do this because its mathematical apparatus is based on the Riemann geometry of time-space, hence prejudicing that 
time-space is there, filled up with fields and matter, that are also there, independent of the observer. There is no fundamental role for
the creation of `new' within relativity theory, everything just `is' and we are only there to `detect' how this everything `is'. That is also
the reason why general relativity can easily be interpreted as delivering a model for the whole universe, whatever this would mean. We know that
quantum mechanics takes into account in an essential way the effect of the observer through the measuring apparatus on the state of the physical
entity under study. In a theory generalizing quantum mechanics and relativity, such that both appear as special cases, this effect should certainly also appear in a fundamental way. 
We believe that general relativity has explored to great depth
the question `how can things {\bf be} in the world'. Quantum axiomatics explores in great depth the question `how can be {\bf acted} in
the world'. And it does explore this question of `action in the world' in a very similar manner as general relativity theory does with its
question of `being of the world'. This means that operational quantum axiomatics can be seen as the development of a general theory of `actions in the world' in the
same manner that Riemann geometry can be seen as a general theory of `geometrical forms existing in the world'. Of course Riemann is not
equivalent to general relativity, a lot of detailed physics had to be known to apply Riemann resulting in general relativity. This is the same
with operational quantum axiomatics, it has the potential to deliver the framework for the theory generalizing quantum mechanics and relativity theory.

We want to remark that in principle a theory that describes the possible actions in the world, and a theory that delivers a model for the whole
universe, should not be incompatible. It should even be so that the theory that delivers a model of the whole universe should incorporate the
theory of actions in the world, which would mean for the situation that exists now, general relativity should contain quantum mechanics, if it
really delivers a model for the whole universe. That is why we believe that Einstein's attitude, trying to incorporate the other forces
and interactions within general relativity, contrary to common believe, was the right one, globally speaking. What Einstein did not know at that
time was `the reality of non-locality in the micro-world'. Non-locality means non-spatiality, which
means that the reality of the micro-world, and hence the reality of the universe as a whole, is not time-space like. Time-space is not the
global theatre of reality, but rather a crystallization and structuration of the macro-world. Time-space has come into existence together with
the macroscopic material entities, and hence it is `their' time and space, but it is not the theatre of the microscopic quantum entities.
This fact is the fundamental reason why general relativity, built on the mathematical geometrical Riemannian structure of time-space, cannot be
the canvas for the new theory to be developed. A way to express this technically would be to say that the set of events cannot be identified with
the set of time-space points as is done in relativity theory. Recourse will have to be taken to a theory that describes reality as a kind of
pre-geometry, and where the geometrical structure arises as a consequence of interactions that collapse into the time-space context. We
believe that operational quantum axiomatics can deliver the framework as well as the methodology to construct and elaborate such a theory.
 
Mackey and Piron introduced the set of yes/no experiments but then immediately shifted to an attempt to axiomatize mathematically the lattice of
(operational) propositions of a quantum entity, Mackey postulating right away an isomorphism with ${\cal L}({\cal H})$ and Piron giving five
axioms to come as close as possible to ${\cal L}({\cal H})$. Also Piron's axioms are however mostly motivated by mimicking mathematically the
structure of ${\cal L}({\cal H})$. In later work Piron made a stronger attempt to found operationally part of the axioms [8], and this
attempt was worked out further in [9], to arrive at a full operational foundation only recently [7]. 

Also mathematically the circle was closed only recently. There
do exist a lot of finite dimensional generalized Hilbert spaces that are different from the three standard examples, real, complex and quaternionic Hilbert space. But since a physical entity
has to have at least a position observable, it follows that the generalized Hilbert space must be infinite dimensional. At the time when Piron gave his five
axioms that lead to the representation within a generalized Hilbert space, there only existed three examples of generalized Hilbert spaces that
fitted all the axioms, namely real, complex and quaternionic Hilbert space. Years later Hans Keller constructed the first counterexample, more specifically an example of an infinite dimensional
generalized Hilbert space that is not isomorphic to one of the three standard Hilbert spaces [10]. The study of generalized Hilbert
spaces, nowadays also called orthomodular spaces, developed into a research subject of its own, and recently Maria Pia Sol\`er proved a
groundbreaking theorem in this field. She proved that an infinite dimensional generalized Hilbert space that contains an orthonormal base is
isomorphic with one of the three standard Hilbert spaces [11]. It has meanwhile also been possible to formulate an operational axiom,
called `plane transitivity' on the set of operational propositions that implies Sol\`er's condition [12], which
completes the axiomatics for standard quantum mechanics by means of six axioms, the original five axioms of Piron and plane transitivity as
sixth axiom.

An interesting and rather recent evolution is taking place, where quantum structures, as developed within this operational approach to quantum axiomatics, are used to model entities in regions of reality different of the micro-world [13, 14, 15, 16, 17, 18, 19, 20]. We believe that also this is a promising evolution in the way to understand deeper and more clearly the meaning of quantum mechanics in all of its aspects.

\bigskip
\noindent
{\it Primary References}

\bigskip
\noindent
[1] Birkhoff, G. and von Neumann, J. (1936), The logic of
quantum mechanics, {\it Annals of Mathematics}, {\bf 37}, 823-843.

\smallskip
\noindent
[5] Mackey, G. (1963), {\it Mathematical Foundations of Quantum Mechanics}, Benjamin,
New York.

\smallskip
\noindent
[6] Piron, C. (1964), Axiomatique quantique, {\it Helvetica Physica Acta}, {\bf 37}, 439-468. 

\smallskip
\noindent
[7] Aerts, D. and Aerts, S. (2004). Towards a general operational and realistic framework for quantum mechanics and relativity theory. In A. C. Elitzur, S. Dolev and N. Kolenda (Eds.), {\it Quo Vadis Quantum Mechanics? Possible Developments in Quantum Theory in the 21st Century} (pp. 153-208). New York: Springer.

\bigskip
\noindent
{\it Secondary References}

\smallskip
\noindent
[2] von Neumann, J. V. (1932), {\it Mathematische Grundlagen der
Quantenmechanik}, Springer, Berlin.

\smallskip
\noindent
[3] Wigner, E.P. (1959), {\it Group Theory and its Applications to Quantum Mechanics of Atomic Spectra}, Academic Press, New York.

\smallskip
\noindent
[4] Mittelstaedt, P. (1963), {\it Philosophische Probleme der Modernen
Physik}, Bibliographisches Institut, Manheim.

\smallskip
\noindent
[8] Piron, C. (1976), {\it Foundations of Quantum Physics}, Benjamin, Massachusetts.

\smallskip
\noindent
[9] Aerts, D. (1982), Description of many physical entities without the paradoxes encountered in quantum mechanics,
{\it Foundations of Physics}, {\bf 12}, 1131-1170.

\smallskip
\noindent
[10] Keller, H. (1980), Ein nicht-klassischer Hilbertscher Raum, {\it Mathematische Zeitschrift}, {\bf 172}, 1432-1823.

\smallskip
\noindent
[11] Sol\`er, M. P. (1995), Characterization of Hilbert spaces by orthomodular spaces, {\it Communications in Algebra}, {\bf 23}, 219-243.

\smallskip
\noindent
[12] Aerts, D. and Van Steirteghem, B. (2000), Quantum axiomatics and a theorem of M. P. Sol\`er, {\it International Journal of Theoretical Physics}, {\bf 39}, 497-502, archive ref and link: quant-ph/0105107. 

\smallskip
\noindent
[13] Eisert, J., Wilkens, M., \& Lewenstein, M. (1999). Quantum games and quantum strategies. {\it Physical Review Letters}, {\bf 83}, 3077-3080.

\smallskip
\noindent
[14] Schaden, M. (2002). Quantum finance: A quantum approach to stock price fluctuations. {\it Physica A}, {\bf 316}, 511-538. 

\smallskip
\noindent
[15] Widdows, D., \& Peters, S. (2003). Word vectors and quantum logic: Experiments with negation and disjunction. In {\it Mathematics of Language 8} (pp. 141-154). Indiana: Bloomington. 

\smallskip
\noindent
[16] Aerts, D., \& Czachor, M. (2004). Quantum aspects of semantic analysis and symbolic artificial intelligence. {\it Journal of Physics A, Mathematical and Theoretical, 37}, L123-L132. 

\smallskip
\noindent
[17] Aerts, D., \& Gabora, L. (2005). A theory of concepts and their combinations I \& II. {\it Kybernetes}, {\bf 34}, 167-191; 192-221.

\smallskip
\noindent
[18] Bruza, P. D. and Cole, R. J. (2005). Quantum logic of semantic space: An exploratory investigation of context effects in practical reasoning. In S. Artemov, H. Barringer, A. S. d'Avila Garcez, L.C. Lamb, J. Woods (Eds.) {\it We Will Show Them: Essays in Honour of Dov Gabbay}. College Publications.

\smallskip
\noindent
[19] Bagarello, F. (2006). An operatorial approach to stock markets. {\it Journal of Physics A}, {\bf 39}, 6823-6840. 

\smallskip
\noindent
[20] Busemeyer, J. R., Wang, Z., \& Townsend, J. T. (2006). Quantum dynamics of human decision making. {\it Journal of Mathematical Psychology}, {\bf 50}, 220-241. 

\end{document}